# PULSAR POPULATIONS AND UNIDENTIFIED γ-RAY SOURCES

A. K. Harding[1], P. L. Gonthier[2], I. A. Grenier[3] and C. A. Perrot[3]

[1]*NASA Goddard Space Flight Center, Greenbelt, MD 20771, USA*
[2]*Department of Physics and Engineering, Hope College, Holland, MI 49423, USA*
[3]*Université Paris VII and CEA/Saclay, Service d'Astrophysique, 91191 Gif-sur-Yvette, France*

## ABSTRACT

The EGRET telescope on the Compton Gamma-Ray Observatory detected over 200 sources and the majority of these are still unidentified. At least three subpopulations of EGRET sources have been associated with the Galaxy: bright sources lying along the Galactic plane, weaker sources spatially correlated with the Gould Belt and a high-latitude, halo population. Many of these sources may be pulsars and there are more than two-dozen radio pulsars in or near EGRET source error boxes, most of them recently discovered in the Parkes Multibeam Survey. We present results from several population synthesis studies of pulsars in the Galaxy, which predict the number of pulsars detected by γ-ray and radio surveys assuming models for the high-energy emission beam and its relation to the radio beam. Future γ-ray pulsar detections by AGILE and GLAST together with the recent large rise in the radio pulsar population will give greatly improved statistics. The relative numbers of radio and γ–ray pulsars detected in the plane and in the Gould Belt populations will provide an important discriminator between models of high-energy emission and will profoundly affect the numbers and birth rate of pulsars in the Galaxy.

## INTRODUCTION

One of the most intriguing mysteries in astronomy is the nature of the unidentified sources of high-energy γ–rays discovered by the EGRET detector on the *Compton Gamma-Ray Observatory* (CGRO). Of these 170 sources (Hartmann et al. 1999), many are at high latitude and are most likely active Galactic nuclei or blazars, but a large majority appears to be Galactic. Numerous multi-wavelength searches (Mukherjee and Halpern 2001) have yielded only a handful of candidate counterparts, primarily because the source positions are determined only to about one degree. Most of the counterparts discovered thus far have turned out to be radio pulsars, a population which makes up the second most numerous class of identified sources in the EGRET catalog, which includes six confirmed and three candidate γ-ray pulsars (Thompson 2001). Many unidentified sources may be radio-quiet γ–ray pulsars, but since EGRET was not able to collect enough photons per source, it was not possible to perform independent period searches to detect these. The anticipated launch of the *Gamma-Ray Large Area Space Telescope* (GLAST) in 2006 is expected to result in the detection of a large number of γ-ray pulsars. Many of these will be radio-selected pulsars, but at least as many may be radio-quiet pulsars (i.e., pulsars undetected by current radio surveys either due to lack of sensitivity or to radio emission not beamed in our direction). GLAST will have the sensitivity to independently detect γ–ray pulsations in all of the EGRET unidentified sources.

The two main types of models proposed to explain pulsar high-energy emission predict different ratios of radio-loud and radio-quiet γ-ray pulsars. Polar cap models (Daugherty and Harding 1996, Rudak and Dyks 1999), where the high-energy and radio emission both originate from the same magnetic polar region, would predict relatively larger overlap between radio and γ-ray pulsar populations than outer gap models (e.g. Romani and Yadigaroglu 1995, Cheng et al. 2000), where the high-energy emission originates from the outer magnetosphere, far from the site of the radio emission. Correspondingly, outer gap models predict much larger numbers of radio-quiet γ–ray pulsars (due to completely missing the radio beam) than do polar cap models. The ratio of radio-loud to radio-quiet γ–ray pulsars will thus be an important discriminator of emission models. We are using population synthesis of the radio and γ-ray pulsars in the Galaxy to predict the number of pulsars that are detectable by current radio surveys and by present and future high-energy γ-ray detectors. Recent results of such simulations will be presented for pulsars in the Galactic plane and in the Gould Belt, a nearby young starburst region.

**UNIDENTIFIED EGRET SOURCES**

There have been a number of statistical studies of the unidentified EGRET sources in an attempt to characterize their spatial distribution and identify possible source subpopulations. One such study (Grenier 2000) identified three Galactic source populations, based on spatial correlations with known Galactic structures. A subpopulation of about 44 sources lies in a thin disk with a scale height of several degrees. Their spectra on average are hard, having photon spectral index of 2.18, they have luminosities of $0.6 - 4 \times 10^{35}$ erg s$^{-1}$ at $1 - 4$ kpc and they have very low flux variability. A second Galactic subpopulation of about 45 sources is correlated with the Gould Belt, a nearby expanding disk of gas and young stars tilted $20^0$ to the Galactic plane. These sources have somewhat softer spectra than those of the plane population, with average photon index of 2.25, and also have fairly low variability. A third subpopulation of roughly 46 sources is distributed in a halo around the galaxy. The halo sources are significantly different from the other two Galactic populations in that they are much more variable and have quite soft average spectra, with photon index 2.52. Another study of the stable unidentified EGRET sources (Gehrels et al. 2000) also found two populations: brighter sources having hard spectra (photon index 2.18) close to the Galactic plane and dimmer sources having softer spectra (index 2.2) correlated with the Gould Belt.

Since the end of the CGRO mission, the Parkes Multibeam Survey (Manchester et al. 2001) has detected many new radio pulsars, currently more than 500. About 25 of these new radio pulsars lie in or near EGRET source error boxes. It has not been possible to verify whether these pulsars are the counterparts of the EGRET sources through epoch-folding of the EGRET archival data to search for γ-ray pulsations, because of pulsar timing noise, possible glitches in the intervening 5-10 years and the small numbers of detected photons. However, it seems likely that young pulsars are good candidates for many of the γ-ray sources near the Galactic plane. Harding and Zhang (2001) suggested that pulsars may also be good candidates for sources in the Gould Belt. Off-beam emission is expected in polar cap models, due to high-altitude curvature radiation from primary electrons. This emission is radiated over a large solid angle, is much softer and has lower luminosity than on-beam emission, characteristics that are very similar to those of the Gould Belt sources. Many would be radio quiet since the line-of-sight will miss the smaller radio beam.

**GALACTIC PLANE PULSARS**

Initial results of a population synthesis study of pulsars in the Galaxy were presented by Gonthier et al. (2002). This study evolves neutron stars from birth distributions in space, magnetic field strength, period and kick velocity, in the Galactic potential to simulate the distribution of radio pulsars detected by eight surveys of the Princeton Catalog. The birth rate of neutron stars is assumed to be constant through the history of the Galaxy and the age of the pulsar is randomly selected from the present to $10^9$ yrs in the past. This initial study assumed a very simple beaming model in which radio and γ-ray beams are aligned with a solid angle of 1 sr. We supply radio and γ-ray characteristics to each neutron star and filter its properties through the set of radio surveys and γ-ray source detection thresholds for EGRET and GLAST. These γ-ray thresholds correspond to the flux required for the instrument to identify the object as a point source, whereas higher flux thresholds are required to detect the pulsation without radio ephemerides. The γ-ray emission model uses the high-energy luminosity from Zhang and Harding (2000), who have modeled the polar cap cascade emission consisting of curvature radiation of primary particles and synchrotron and inverse-Compton radiation of pairs. This model uses the self-consistent space-charge limited flow acceleration of Harding and Muslimov (1998) to energize the primary particles. Gonthier et al. (2002) found that agreement of the $\dot{P} - P$ distribution of simulated radio pulsars with the observed distribution was significantly improved by assuming decay of the neutron star surface magnetic field on a timescale of 5 Myr. In obtaining this result, however, they ignored the period dependence of the beaming. In this case, EGRET should have detected 9 radio-loud and 2 radio-quiet γ-ray pulsars, and GLAST should detect 90 radio-loud and 101 radio-quiet pulsars (9 detected as pulsed sources). Because the radio and γ-ray beam apertures were assumed to be identical, "radio-quiet" γ-ray pulsars are those whose radio emission is too weak to be detected by the selected radio surveys.

Following this initial study, we have made several improvements to the Monte Carlo simulation. The addition of Parkes MB Survey pulsars has doubled the sample of detected radio pulsars from 445 to presently 906 (which will increase further as the survey nears completion and more pulsars are timed). We have used the recently improved distance model of Cordes and Lazio (2002) to determine the dispersion measure of the simulated pulsars. Also, the treatment of the radio and γ-ray beam geometry has been substantially improved. Each beam is modeled independently but with a common axis of symmetry, the magnetic axis. The radio beam is based on an empirical model of Arzoumanian, Chernoff and Cordes (2002), who define a flux, $s(\theta)$ as a function of angle $\theta$ to the magnetic axis, composed of a core component along the magnetic axis and a conal component along the edge of a hollow cone defined by the last open field line,

$$s(\theta) = s_{core} e^{-\theta^2/\rho_{core}^2} + s_{cone} e^{-(\theta-\bar{\theta})^2/\rho_{cone}^2} \quad (1)$$

where $\rho_{core} = 1.5^0 P^{-0.5}$ and $\rho_{cone} = 1.4^0 (1+66/\nu_{obs}) P^{-0.5}$ are the core and cone widths at observing frequency $\nu_{obs}$ and $\bar{\theta} = 3.9^0 P^{-0.5}$ is the cone radius. The total radio luminosity is $L_R = 10^{29.3} P^{-1.5} \dot{P}^{0.4}$ erg s$^{-1}$, with the ratio of core to cone luminosity being $(20/3P)(400 \text{ MHz}/\nu_{obs})^{0.5}$. The γ-ray luminosity and beam geometry are based on a theoretical model of emission and acceleration in a slot gap near the rim of the polar cap. This model, originally developed by Arons and Scharlemann (1979) and Arons (1983), has been revised by Muslimov and Harding (2003) to include particle acceleration due to inertial frame-dragging (Muslimov and Tsygan 1992). In this model, particles at the outer rim of the polar cap are accelerated relatively slowly from the neutron star surface to altitudes of several stellar radii, where they produce synchrotron-pair cascades.

A hollow cone of on-beam emission results from synchrotron radiation of pairs and a very broad cone of off-beam emission results from curvature radiation of primaries above the pair-cascade zone. Off-beam γ-ray emission is thus visible at large angles to the magnetic axis. We have developed a simplified, analytic description of the γ-ray beam based on this model. Figure 1 shows schematically the radio and γ-ray beam geometry assumed. In this scheme, the radio core beam would lie in phase between the two γ-ray conal peaks in the pulse profile. The fact that the radio peak often leads the γ-ray peaks in the pulse profiles of the known γ-ray pulsars indicates that polar-cap γ-ray emission and radio core-dominance in young (short-period) pulsars are not compatible.

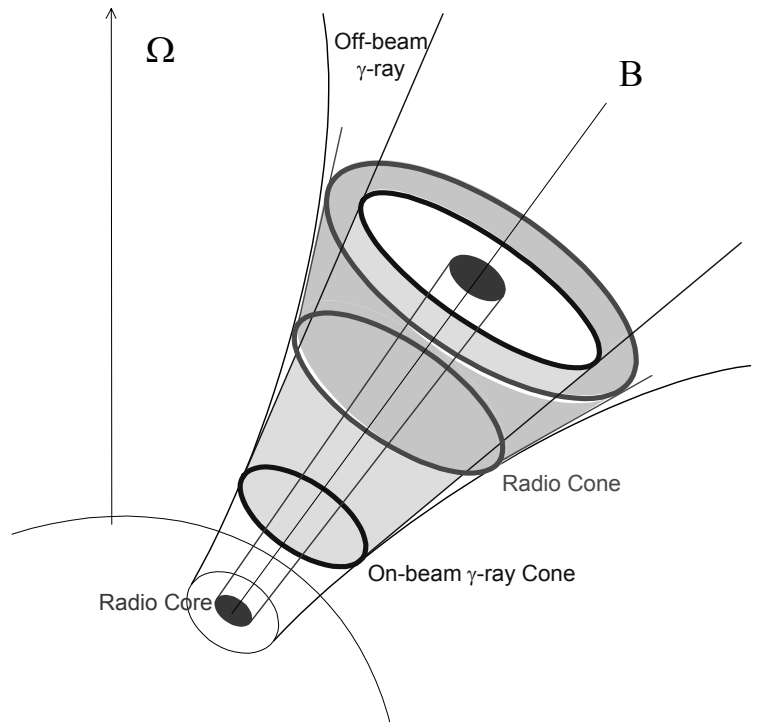

Fig. 1. Geometry of polar cap radio and γ-ray beams

Preliminary results are shown in Figure 2, which displays a comparison of $\dot{P} - P$ plots of observed and simulated radio and γ-ray pulsars. The left panel shows radio pulsars detected by nine surveys (including Parkes MB) and γ-ray pulsars detected by EGRET. The right panel shows the radio and γ-ray pulsars "detected" in the simulation by the radio surveys and by EGRET. The Parkes MB survey detected many more radio pulsars with magnetic fields above $10^{13}$ G than previous surveys. We find as a result that a broader magnetic field birth distribution and a shorter decay timescale is required in order to populate the high-field region of the $\dot{P} - P$ plot without producing too many high-field pulsars with long periods. The simulation shown in Figure 2 includes magnetic-field decay on a 2.8 Myr timescale. Thus, there is an indication that higher fields decay on shorter timescales, a result that is not unexpected on theoretical grounds (e.g. Goldreich and Reisnegger 1992, Heyl and Kulkarni 1998). The simulation in this case results in 19 radio-loud and 4 radio-quiet pulsars detected by EGRET and 324 radio-loud and 207 radio-quiet pulsars detected by GLAST, with a derived neutron star birth rate of 1.1 per century. In comparison, a study simulating γ-ray pulsars in the outer gap model (Zhang et al. 2000) and modeling radio pulsars from the Princeton catalog predicts 10 radio-loud and 22 radio-quiet pulsars detectable by EGRET and 80 radio-loud and 1100 radio-quiet pulsars detectable by GLAST. The much larger ratio of radio-loud to radio-quiet pulsars predicted for both EGRET and GLAST by the outer gap model is due to the outer-gap γ-ray beam being much larger than the radio beam.

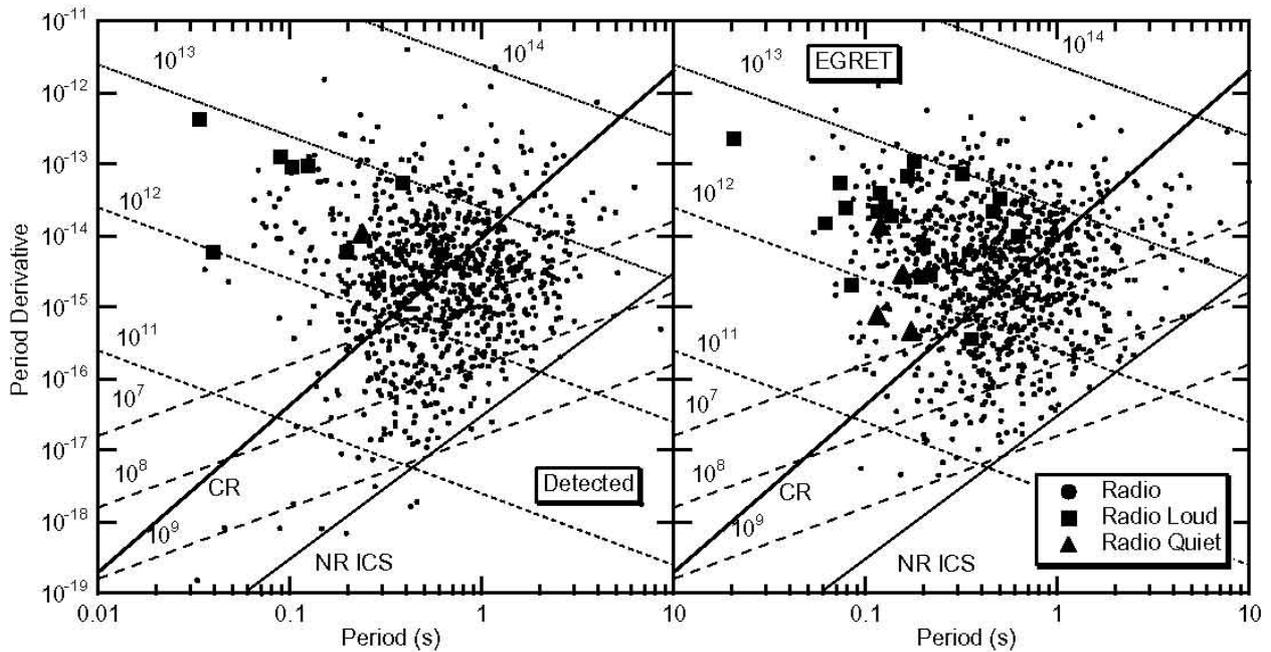

Fig. 2. Period derivative - period distributions of observed pulsars (left panel) and simulated pulsars (right panel) assuming field decay. Small solid dots indicate detected radio pulsars, solid squares represent radio-loud γ-ray pulsars, and solid triangles are radio-quiet γ-ray pulsars detected (left) and predicted (right) for EGRET.

The addition of the Parkes MB survey pulsars more than doubles the number of radio pulsars in the sample and it increases our predicted number of EGRET radio-loud γ-ray pulsars from 9 to 19. Thus many of the unidentified EGRET sources in the galactic plane may be young, radio-loud γ-ray pulsars. Many of these newly discovered young and energetic radio pulsars are too distant to have been discovered in previous radio surveys, which were limited in their ability to find sources of pulsed emission that are highly dispersed and scattered by the high electron density in the plane. Gamma-ray telescopes may therefore have an advantage in detecting young and distant neutron stars in the galaxy.

**GOULD BELT PULSARS**

The Gould Belt is a nearby (200 – 300 pc) disk of young stars and gas surrounding the Sun. The 30 – 40 Myr old system has a supernova rate estimated to be 3-5 times higher than that in the local Galactic plane (Grenier 2000) and is inclined at an angle of about $20^0$ to the plane. Since one subpopulation of EGRET unidentified sources is found to be spatially correlated with the Gould Belt, it is possible that its enhanced supernova rate (and thus birthrate of neutron stars) may point to the nature of these sources. We have begun a population synthesis study of pulsars in the Gould Belt, evolving neutron stars born in the Belt from their birthplaces. Recently, Perrot and Grenier (2002) have modeled the 3D dynamical evolution of the gas in the Gould Belt over the last 30 Myr. This model has been used to simulate the present population of pulsars in the Belt, their γ-ray emission properties and their detectability, by EGRET in order to determine how many of the unidentified γ-ray sources are likely to be pulsars. The pulsar population in the Galactic plane is simultaneously simulated in a way similar to that described in Gonthier et al. (2002). An important difference is that this study assumes a birth rate of neutron stars in the Gould Belt, based on a supernova rate of 20 $Myr^{-1}$, in the galaxy using a supernova rate of 2.13 per century.

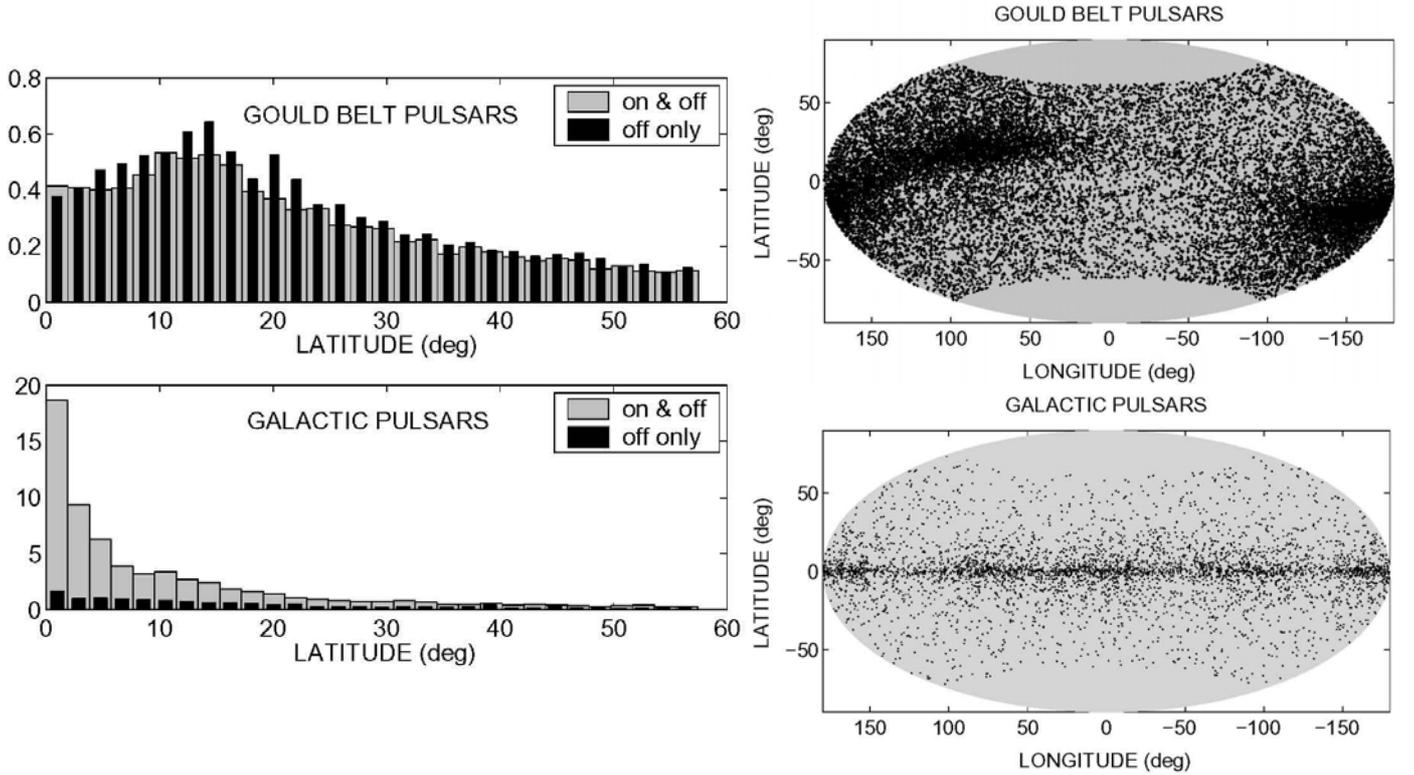

Fig. 3. Spatial distributions of simulated on-beam and off-beam γ-ray pulsars in the Gould Belt and the galactic plane. The vertical scale of the left-hand plots is number density (top) and number (bottom). For the "on & off" pulsars both on-beam and off-beam emission cones are seen, while for the "off only" pulsars, only the off-beam cone is visible (see Figure 1). The sky plots on the right show both cases combined

The neutron stars are evolved only over 5 Myr. The pulsar emission model assumes on-beam and off-beam γ-ray emission cones with opening angle, $\theta_{ON}, \theta_{OFF} \propto P^{-1/2}$, centered on the magnetic axis, at random inclination to the spin axis (see Figure 1). The emission cones have luminosities, $L_{ON} = \lambda_{ON} \dot{E}^{1/2}$ and $L_{OFF} = \lambda_{OFF} \dot{E}^{1/2}$, reflecting the correlation of γ-ray luminosity with spin-down power of the known γ-ray pulsars (Thompson 2001). The luminosity and opening angle of the on-beam and off-beam cones are free parameters determined by a maximum likelihood fit to the detected populations of EGRET γ-ray sources in the plane and in the Gould Belt.

Fitting the 104 persistent γ-ray sources found at |b|<60°, we find 79 (16 off-beam and 63 on-beam) sources in the plane and 19 (10 off-beam and 9 on-beam) sources in the Belt. The simulated Gould-Belt sources have an average age of 1.5 Myr, significantly older than the Galactic-plane sources that have average age 0.35 Myr, as expected for more distant sources. Although the total beam luminosity of the pulsars producing the on-beam and off-beam emission in the Gould Belt is in the same range of $L_{ON} \sim L_{OFF} \sim 10^{32} - 10^{34}$ erg s$^{-1}$, the emission solid angles, $\Omega_{on} < 0.5$ sr. and $\Omega_{off} \sim 2\pi$ sr., are very different, so that the flux of the off-beam sources is predicted to be at least a factor of $4\pi$ smaller than that of the on-beam sources. However, all of the on-beam pulsars and most of the off-beam pulsars in the Gould Belt are detectable by EGRET. To be detectable in the Galactic plane, on-beam pulsars must be closer than 4 kpc and off-beam pulsars closer than 1 kpc because of the reduced EGRET sensitivity at low latitude.

Figure 3 shows the resulting spatial distributions of Gould Belt and Galactic-plane pulsars. The signature of the Gould Belt, a mid-latitude band crossing the Galactic plane, is preserved over 5 Myr, even though many of the pulsars born with high velocities have escaped the Belt. Thus, one would expect the detectable on-beam and off beam γ-ray pulsars born in the Belt to trace its structure. Note that relatively few off-beam pulsars in the Galactic plane population were detectable by EGRET, as radio-quiet pulsars.

## CONCLUSIONS

Population synthesis studies of pulsars born in the Galactic plane and in the Gould Belt are providing estimates of the number of radio-loud and radio-quiet γ-ray pulsars in the Galaxy that are detectable by both past and future γ-ray detectors. These estimates are necessarily model dependent, but they consequently can be a powerful tool for discriminating between the different possible models for pulsar high-energy emission. They can also help to solve the mystery of the unidentified EGRET sources and be an important guide to future missions such as AGILE and GLAST in their observation of these sources. Combined synthesis of radio and γ-ray pulsars in the Galaxy using a polar-cap model geometry indicates that EGRET should have detected many more radio-loud than radio-quiet γ-ray pulsars. But GLAST is predicted to detect a higher ratio of radio-quiet to radio-loud pulsars (although still more radio-loud pulsars). Comparison of observed and simulated radio pulsar distributions improves if the surface magnetic field is assumed to decay on a several million year timescale. Although the actual numbers are sensitive to the parameters assumed in the models, it is very clear that outer gap models predict a much higher ratio of radio-quiet to radio-loud γ-ray pulsars than do polar cap models for both EGRET and GLAST. These results also imply that outer gap models predict a much greater fraction of radio-quiet to radio-loud pulsars among the EGRET unidentified sources in the plane than do polar cap models. The results of searches by AGILE and GLAST for the γ-ray pulsar counterparts to the new radio pulsars detected in EGRET error boxes will greatly help in constraining models. Simulations of pulsars born in the Gould Belt show that the γ-ray pulsars retain the spatial signature of the Belt over 5 Myr. About 50% of the EGRET sources associated with the Gould Belt could be γ-ray pulsars, with about equal numbers of on-beam (bright emission near the magnetic axis) and off-beam (fainter emission further from the magnetic axis) sources visible.

E-mail address of A. K. Harding  Alice.K.Harding@nasa.gov